\documentclass[mathleft
]{an}
\usepackage{graphicx}
\usepackage{times}
\begin{document}
\clubpenalty10000
\sloppy
\def\kcet{$\kappa^1$~Ceti } 
% The following seven commands are intended for editorial usage and should be ignored by
% the author(s).
\Pagespan{000}{000}% Document's page range. 
% If second parameter is left empty, the last page is computed automatically.
\Yearpublication{2010}%
\Yearsubmission{2010}%
\Month{00}%   
\Volume{0000}%  
\Issue{00}% 
% \DOI{This.is/not.aDOI}% 

\title{How did the Sun affect the climate when life evolved on the Earth? -- A case study on the young solar twin \kcet}

\author{C. Karoff\inst{1,2}\fnmsep\thanks{\email{karoff@bison.ph.bham.ac.uk}} \& H. Svensmark\inst{3}\newline
%Example 
%for footnote, note the usage of the \texttt{fnmsep}
%command as separator between institute number and footnote mark} 
}
\titlerunning{How did the Sun affect the climate when life evolved on the Earth?}
\authorrunning{Karoff \& Svensmark}
\institute{
School of Physics and Astronomy, University of Birmingham, 
Edgbaston, Birmingham B15 2TT, UK
\and
Department of Physics and Astronomy, Aarhus University, Ny Munkegade 120, DK-8000 Aarhus C, Denmark
\and
Danish National Space Center, Danish Technical University, Juliane Maries Vej 30,  DK-2100 Copenhagen {\O}, Denmark
}

\received{??}
\accepted{??}
\publonline{later}


\abstract{Using \kcet as a proxy for the young Sun we show that not only was the young Sun much more effective in protecting the Earth environment from galactic cosmic rays than the present day Sun; it also had flare and corona mass ejection rates up to three orders of magnitude larger than the present day Sun. The reduction in the galactic cosmic ray influx caused by the young Sun's enhanced shielding capability has been suggested as a solution to what is known as the faint young Sun paradox, i.e. the fact that the luminosity of the young Sun was only around 75\% of its present value when life started to evolve on our planet around four billion years ago. This suggestion relies on the hypothesis that the changing solar activity results in a changing influx of galactic cosmic rays to the Earth, which results in a changing low-altitude cloud coverage and thus a changing climate. Here we show how the larger corona mass ejection rates of the young Sun would have had an effect on the climate with a magnitude similar to the enhanced shielding capability of the young Sun.}

\maketitle

\section{Introduction}
Stellar evolution models predict that when life evolved on the Earth more the 4 billion years ago just after the heavy bombardment, the conditions in the inner part of the solar system were significantly different compared to what we know today. E.g. it was noted already in 1972 by Sagan and Mullen that the luminosity of the young Sun was only around 75\% of its present value, which would result in freezing temperatures on the Earth --  assuming a radiation budget similar to what we have today (Sagan \& Mullen 1972, see also Gough 1981). This problem has since been known as {\it the faint young Sun paradox}. On the other hand stellar evolution models predict that the activity of the Sun related to the chromosphere and corona should be much stronger than what we know today. 

A number of different solutions have been proposed to {\it the faint young Sun paradox}. Sagan \& Mullen (1972) suggested that elevated levels of CO$_{2}$ could have maintained surface temperatures above freezing, but the CO$_{2}$ level needed to raise the temperature might be so high that it would be in conflict with geochemical records (Rye et al. 1995). Potentially other greenhouse gasses like ammonia could also help to raise the temperature, but on the other hand it is questionable if large amounts of ammonia could be maintained in the Earth atmosphere at a time where the Sun's UV radiation was up to 10 times larger than today (Sagan \& Chyba 1972).  

Another solution that was originally proposed by Graedel (1991) suggests that the Sun has experienced a mass loss of 5-10\% over its main-sequence life. Therefore the young Sun would have been a bit more massive than today's Sun and thus brighter than what we would expect without considering mass loss. Unfortunately mass loss rates as high as 10\% contradicts both solar evolution models calibrated using helioseismology (Guzik \& Cox 1995) and measurements of stellar winds around solar-type stars (Gaidos et al. 2000; Wood et al. 2002).

Recently attempts to solve {\it the faint young Sun paradox} have been based on atmosphere models including not only an increased greenhouse effect, but also a reduced albedo (von Paris et al. 2008; Kitzmann et al. 2010). These attempts are in line with the hypothesis put forward by Svensmark \& Friis-Christensen (1997): that galactic cosmic rays (GCRs) modulate the amount of aerosols and clouds in the lower part of the Earth's atmosphere. In other words, in order to understand the evolution of the Earth's climate now and back then, it is important to understand not only the effect from high-altitude clouds (through the greenhouse effect), but also the effect from low-altitude clouds (through the albedo effect). By including both effects in there modes it was suggested by Shaviv (2003) and Svensmark (2003, 2006) that the young active Sun's increased ability to protect the Earth from GCRs could cause higher temperatures on the Earth. 

The hypothesis is that as the Sun was much more active when life started to evolve on the Earth (which is reflected in its higher rotation rate and higher level of UV and X-ray emission) it would have been much more efficient in shielding us from the GCRs, which would have resulted in smaller amounts of aerosols and clouds in the lower part of the Earth's atmosphere and thus higher temperatures on the Earth. This hypothesis was recently strengthened by observations of direct evidence of a relation between GCRs, aerosols and clouds on short time scales as observed during major Forbush decreases (Svensmark, Bondo \& Svensmark 2009).

Though it was noted by Svensmark, Bondo \& Svensmark (2009) that large Forbush decreases today are too rare to have any significant effect on the Earth's climate, this might not have been the case 4 billion years ago when life started to evolve on the Earth as CMEs are expected to have been much more common on the early Sun. We therefore analyze if the reduction in the influx of GCRs originating from Forbush decreases could be significant compared to the reduction in the influx of GCRs originating from the more effective shielding capacity of Sun at the time life evolved on the Earth. 

We undertake this analysis through a case study of the young solar twin \kcet. With an age of around 700 million years \kcet mimics the Sun at that time. The age estimate of \kcet is based on the rapid rotation of \kcet with a period of 8.6 days (Rucinski et al. 2004) and a resulting large activity level (Baliunas et al. 1995), but the uncertainties of such a simple scaling relation for the ages are of course huge. Other fundamental stellar parameters of \kcet such as effective temperature, surface gravity and metallicity come so close to solar values that \kcet qualify as a solar analogue (see Table~1). By using the well studied \kcet as a case study instead of (simple) scaling laws between stellar activity and age [as done by Shaviv (2003) and Svensmark (2003, 2006)] we can base our analysis on actual measurements of stellar activity rather than estimates based on physical assumptions.  

\begin{table}
\caption{Stellar parameters for {$\kappa^1$} Ceti (from Gaidos \& Gonzalez 2002) and the Sun (from Christensen-Dalsgaard et al. 1996)}
\centering
\begin{tabular}{lccccc}
\hline \hline
Name & Type & $T_{\rm eff}$ [K] & log $g$ & [Fe/H] \\
\hline
{$\kappa^1$} Ceti & G5 V &  5747 (49) & 4.53 (0.06) & 0.11 (0.04)\\
Sun & G2 V &  5778 & 4.44 & 0.00 \\
\hline
\end{tabular}
\label{tab1}
\end{table}

\section{The effect of more effective shielding}
The temperature response to a change in the GCR influx is given by (Shaviv, 2003):

\begin{equation}
\Delta T_{GCR} \approx D[1-(\varepsilon_{\rm \kappa^1 Ceti}/\varepsilon_{\odot})^q],
\end{equation}

where $\varepsilon_{\rm \kappa^1 Ceti}$ is the GCR flux reaching the troposphere around an Earth-like planet around \kcet and $\varepsilon_{\odot}$ is the GCR flux reaching the troposphere around the Earth. $D$ and $q$ are constants ($D$ $\sim$ 10 K and $q$ $\sim$ 0.5).

The GCR influx can be found by solving the spherically-symmetric transport equation for the stellar modulation of cosmic rays reaching the planet's troposphere (Perko, 1987):

\begin{equation}
\frac{\partial U}{\partial r} +\frac{VP}{3 \kappa} \frac{\partial U}{\partial P} \simeq 0,
\end{equation}
where $U$ is the cosmic ray distribution function, $r$ is the heliocentric radial distance, $P$ is the particle rigidity, $V$ is the solar wind speed and $\kappa$ is the diffusion coefficient for radial propagation. The spherically-symmetric transport equation can be solved for GCRs with energies larger than a few GeV using the force-field approximation. In the force-field approximation the kinetic energies of the GCRs at the planet $E$ is given as:

\begin{equation}
E=E_{\rm ISM} - \Phi,
\end{equation} 
where $E_{\rm ISM}$ is the kinetic energy of the GCRs in the interstellar medium at the astrospheric boundary and $\Phi$ is the modulation strength (Perko, 1987):

\begin{equation} 
\Phi = \frac{rV}{3\kappa}.
\end{equation}
If we assume that the stellar wind speed and the diffusion coefficient is the same for the Sun and \kcet the modulation strength for \kcet can be found by assuming that the ram pressure of the wind around \kcet equals that of the interstellar medium and therefore is the same as for the Sun:

\begin{equation}
P_{\rm ram} = \rho V^2 \propto \frac{\dot{M}V}{r^2},
\end{equation}
so 
\begin{equation}
\Phi \propto r \propto \sqrt{\dot{M}}.
\end{equation}
The mass loss of \kcet has been measured by Gaidos (1998) and Gaidos, G{\"u}del \& Blake (2000) to $\dot{M}_{\rm \kappa^1 Ceti}\sim4\cdot 10^{-11} {\rm M_{\odot}/yr}$. 

Following Shaviv (2003) we can now calculate the energy of the GCR influx to an Earth-like planet around \kcet relative to the energy of the GCR influx to the Earth:

\begin{equation}
\frac{\varepsilon_{\rm \kappa^1 Ceti}}{\varepsilon_{\odot}} = \frac{\int_{E_{\rm c}}^{\infty} f_{\rm \kappa^1 Ceti}E_{\rm \kappa^1 Ceti}dE}{\int_{E_{\rm c}}^{\infty} f_{\odot}E_{\odot}dE},
\end{equation}
where $E_{\rm c}$ is the cutoff energy of GCRs that can actually reach the troposphere ($\sim$ 12 GeV) and $f$ is the differential number flux reaching an Earth-like planet around either the Sun or \kcet which again is a function of the differential number flux of the interstellar medium (Shaviv, 2003):

\begin{equation}
f \propto (E+\Phi)^{-2.7}.
\end{equation}
We thus obtain:

\begin{equation}
\frac{\varepsilon_{\rm \kappa^1 Ceti}}{\varepsilon_{\odot} }=
\frac{\left(E_{\rm c} +\Phi_{\rm \kappa^1 Ceti} \right)^{-1.7}}  { \left(E_{\rm c} +\Phi_{\odot} \right)^{-1.7} }
\frac{\left( \frac{E_{\rm c}}{0.7} +\frac {\Phi_{\rm \kappa^1 Ceti}}{1.7} \right)}{\left(\frac{E_{\rm c}}{0.7} \frac{\Phi_{\odot}}{1.7} \right) },
\end{equation}
or $\varepsilon_{\rm \kappa^1 Ceti}/\varepsilon_{\odot} \sim 0.1$ -- i.e. an Earth-like planet around \kcet would only receive around 10\% of the cosmic ray flux that we receive on the Earth and the temperature would thus be around 7 degrees warmer than it would have been had \kcet not been able to protect its planet more effectively from GCRs than the Sun.

\section{The effect of Forbush Decreases}
CMEs directed toward the Earth can lead to sudden reductions in the influx of GCRs over time scales from hours to days known as Forbush decreases. It was shown by Svensmark, Bondo \& Svensmark (2009) that large Forbush decreases were followed by reduced levels of aerosols, of cloud water content, of liquid water cloud fraction and of low IR-detected clouds. 

Svensmark, Bondo \& Svensmark (2009) analyzed five CMEs found over a time span of 10 years which all resulted in Forbush decreases associated with an  $\sim$10 \% decrease in the GCR influx over around a week - this led them to note that large Forbush decreases today are too rare to have any significant effect on the Earth's climate. On the other hand it is not given that large Forbush decreases did not have a significant effect on the Earth's climate 4 billion years ago when life started to evolve on the Earth, as CMEs are expected to have been much more common on the early Sun. We therefore analyze if the reduction in the influx of GCRs originating from Forbush decreases could be significant compared to the reduction in the influx of GCRs from the more effective shielding from the Sun at the time life evolved on the Earth. 

Assuming that the CME rate scales liniarly with the flare rate (which seems to be the case for the Sun) we can use the flare rate of \kcet to provide us with an estimate of the CME rate and thus an estimate of how common Forbush decreases would be on an Earth-like planet  around $\kappa^1$~Ceti. The cumulative flare occurrence rate distribution of \kcet was calculated by Audard et al. (2000) using 7 days of EUV observations from the {\it Extreme Ultraviolet Explorer} (Malina \& Bowyer 1991). In order to compare this cumulative flare occurrence rate distribution to the Sun we have analyzed the occurrence of solar flares from 1998 to 2007 observed in X-ray by the {\it Geostationary Operational Environmental Satellite} system\footnote{The {\it Geostationary Operational Environmental Satellite} system observations were obtained from http://www.ngdc.noaa.gov/stp/SOLAR/ftpsolarflares.html} (Garcia 1994).

The flare occurrence rate distribution for the Sun and \kcet are shown in Fig.~1 (solid lines) together with power law fits to these distributions of the form (Audard et al. 2000): $N(>E)=kE^{-\alpha+1}$,
where $k$ is a normalization factor and $\alpha$ is a constant measuring the hardness of the flare distribution.

In order to compare the two flare occurrence rate distributions we need to correct them for the fact that the solar flares were observed in an energy range roughly 50 times smaller than energy range in which the flares on \kcet where observed in; the solar flares were observed in the soft X-ray band (1-8{\AA), while the flares on \kcet were observed in the EUV band (0.01-10 {\rm KeV}). We therefore multiply the solar flare energies with 50 (Audard et al. 2000).

\begin{figure}
\includegraphics[width=\columnwidth]{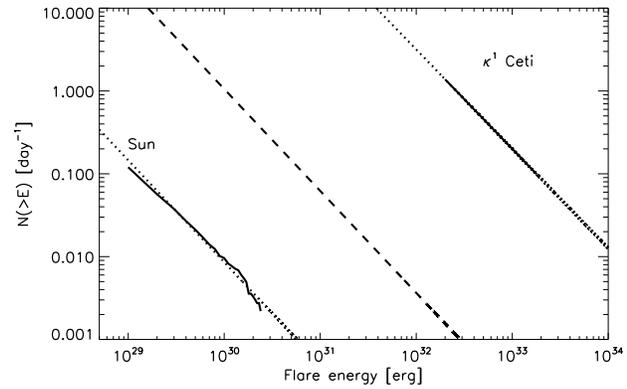}
\caption{Comparison between the cumulative flare occurrence rate distributions for the Sun and $\kappa^1$~Ceti. The flare rate for the Sun has been calculated from soft X-ray (1-8 {\AA}) data from {\it The Geostationary Operational Environmental Satellite} system, and the flare rate for \kcet has been calculated from EUV data from the  {\it Extreme Ultraviolet Explorer} (0.01-10 {\rm KeV}) from Audard et al. (2000). The solid lines show the observations. The dotted lines show power law fits to the observations of the Sun and \kcet, respectivily. The dashed line shows the power law fit to the solar observation, but here the flare energies have been multiplied by 50 in order to make a reliable comparison to the \kcet observations, which have been integrated over a larger energy range. It is seen that whereas flares with integrated energies around $10^{32}$ erg (the ones that causes Forbush decreases) are rather rare on the Sun, they occure daily on $\kappa^1$~Ceti.} 
\end{figure}
All the five Forbush decreases analyzed by Svensmark, Bondo \& Svensmark (2009) were associated with flares with an integrated energy around $10^{32}$ erg. These Forbush decreases generally led to a $\sim$10 \% decrease in the GCR influx over around a week. In Fig.~1 it is seen that such flares happen around once a day on $\kappa^1$~Ceti. This means that around 7 Forbush decreases would be present around an Earth-like planet around \kcet at any given time. Assuming that all 7 Forbush decreases will lead to a 10\% reduction in the GCR influx, an Earth-like planet around \kcet would be experiencing an approximate 50\% mean reduction in the GCR influx from Forbush decreases -- i.e around half the reduction that is expected to occur due to the more effective shielding of GCR around $\kappa^1$~Ceti.

It is apparent that an Earth-like planet around \kcet would be experiencing a reduction in the GCR influx from both a more effective shielding from a larger astrosphere and from Forbush decreases at the same time. Thus it would experience a 90\% reduction from more effective shielding from a larger astrosphere and a 50\% reduction of the remaining 10\% from Forbush decreases. Therefore by adding the contribute from Forbush decreases to the contribute from more effective shielding from a larger astrosphere Eq.~1 predicts an 8 instead of a 7 degree warmer climate. This is of course an insignificant difference -- i.e. it does not change much to remove what is absent. On the other hand this study has shown that if the early Sun had not been more effective in shielding the Earth from GCRs, the Earth would still have experienced a reduced GCR influx due to Forbush decreases. 

\section{Conclusion}
Using the young solar twin \kcet as a case study we have shown that the reduction in the GCR influx to the Earth caused by Forbush decreases had the same order of magnitude as the reduction caused by more effective shielding from a larger heliosphere at the time life evolved on the Earth.

This does not change the conclusion made by Shaviv (2003): that the warming associated with a reduced GCR influx is enough to significantly compensate for the fainter Sun at the time life evolved on the Earth and can explain about 1/2 to 2/3's of the temperature increase between now and then. Thus the warming associated with a reduced GCR influx is enough to solve {\it the faint young Sun paradox}.

An open question is whether the GCRs that are scattered away from the CMEs in the Forbush decrease will eventually return. There is no evidence from ground-based observations of GCRs that the influx increases after a Forbush decrease and it therefore seems secure to assume that the 10\% of the GCR influx is simply scattered so much away from the Earth during a Forbush decrease that it can be considered removed from the near-Earth environment. This is most likely also the case for an Earth-like planet around \kcet as a large part of the CMEs will be magnitudes larger than what we have observed on the Sun.

\kcet is a unique laboratory for understanding the activity of the Sun when life evolved on the Earth. Not only for understanding how a reduced GCR influx could affect the climate back then, but also for understanding for example how a larger X-ray flux from the larger corona and a larger UV flux from a stronger chromosphere could affect the evolution of life. 

The larger flare rate of the young Sun has of course also had other consequences. Firstly, the largest flares and accompanying CMEs would have reduced the amount of ozone in the Earth atmosphere, making life on Earth much more vulnerable to UV radiation (Schaefer et al. 2000). Secondly, large solar flares and accompanying CMEs might also have played a more direct r\^{o}le in the evolution of life on the Earth by providing an energy source to create organic molecules such as the lightning in
the Miller-Urey experiment (Miller \& Urey 1959).

Unfortunately, we still lack a good asteroseismic estimate of the age of this star. This is unfortunate because the best current age estimates of \kcet based on measurements of the rotation and activity of \kcet come with uncertainties of $\sim$ 500 million years. \kcet is therefore an obvious target for future ground-based asteroseismic campaigns -- e.g. the first observations by the {\it Stellar Observations Network Group} (Grundahl et al. 2009).

\vspace{2cm}
\section*{Acknowledgments}
CK acknowledges financial support from the Danish Natural Science Research Council. The National Oceanic and Atmospheric Administration operates the {\it Geostationary Operational Environmental Satellite} system.

\end{document}